\documentstyle[12pt]{article}
\catcode`\@=11
\def\marginnote#1{}
\newcount\hour
\newcount\minute
\newtoks\amorpm
\hour=\time\divide\hour by60
\minute=\time{\multiply\hour by60 \global\advance\minute by-\hour}
\edef\standardtime{{\ifnum\hour<12 \global\amorpm={am}%
        \else\global\amorpm={pm}\advance\hour by-12 \fi
        \ifnum\hour=0 \hour=12 \fi
        \number\hour:\ifnum\minute<10 0\fi\number\minute\the\amorpm}}
\edef\militarytime{\number\hour:\ifnum\minute<10 0\fi\number\minute}
\def\draftlabel#1{{\@bsphack\if@filesw {\let\thepage\relax
   \xdef\@gtempa{\write\@auxout{\string
      \newlabel{#1}{{\@currentlabel}{\thepage}}}}}\@gtempa
   \if@nobreak \ifvmode\nobreak\fi\fi\fi\@esphack}
        \gdef\@eqnlabel{#1}}
\def\@eqnlabel{}
\def\@vacuum{}
\def\draftmarginnote#1{\marginpar{\raggedright\scriptsize\tt#1}}
\def\draft{\oddsidemargin -.5truein
        \def\@oddfoot{\sl preliminary draft \hfil
        \rm\thepage\hfil\sl\today\quad\militarytime}
           \let\@evenfoot\@oddfoot \overfullrule 3pt
        \let\label=\draftlabel
        \let\marginnote=\draftmarginnote
   \def\@eqnnum{(\theequation)\rlap{\kern\marginparsep\tt\@eqnlabel}%
\global\let\@eqnlabel\@vacuum}  }

\def\preprint{\twocolumn\sloppy\flushbottom\parindent 1em
        \leftmargini 2em\leftmarginv .5em\leftmarginvi .5em
        \oddsidemargin -.5in    \evensidemargin -.5in
        \columnsep 15mm \footheight 0pt
        \textwidth 250mmin      \topmargin  -.4in
        \headheight 12pt \topskip .4in
        \textheight 175mm
        \footskip 0pt
        \def\@oddhead{\thepage\hfil\addtocounter{page}{1}\thepage}
        \let\@evenhead\@oddhead \def\@oddfoot{} \def\@evenfoot{} }
\def\titlepage{\@restonecolfalse\if@twocolumn\@restonecoltrue\onecolumn
     \else \newpage \fi \thispagestyle{empty}\c@page\z@
        \def\thefootnote{\fnsymbol{footnote}} }

\def\endtitlepage{\if@restonecol\twocolumn \else  \fi
        \def\thefootnote{\arabic{footnote}}
        \setcounter{footnote}{0}}  %\c@footnote\z@ }
\catcode`@=12
\relax
\def\bea{\begin{array}}
\def\bem{\begin{displaymath}}
\def\beq{\begin{equation}}
\def\eea{\end{array}}
\def\eem{\end{displaymath}}
\def\eeq{\end{equation}}

\def\NP#1#2#3{Nucl. Phys. \underline{#1} (19#2) #3}

\def\PL#1#2#3{Phys. Lett. \underline{#1} (19#2) #3}

\relax
\newcommand{\be}{\begin{equation}}
\newcommand{\en}{\end{equation}}
\newcommand{\ba}{\begin{eqnarray}}
\newcommand{\ea}{\end{eqnarray}}
\newcommand{\ee}{\end{equation}}

\relax
\begin{document}
\topmargin-.4cm
\renewcommand{\theequation}{\thesection.\arabic{equation}}
%\draft
%\preprint
\begin{titlepage}
\begin{flushright}
LPTENS-00/48\\
hep--th/0012192\\
December 2000
\end{flushright}
\vspace{1cm}
\begin{center}{\Large\bf
Type II NS Five-Branes:\\
\vskip .2cm
Non Critical Strings \& their Topological Sectors$^{\star}$ }
\vskip 1cm
{\large \bf  Costas Kounnas}\\
\vskip .5cm
{\normalsize\sl
{Laboratoire de Physique Th\'eorique,
ENS, F-75231 Paris France $\, ^{\dagger}$}}
\end{center}
\vskip .5cm
\begin{center}
{\bf Abstract}
\end{center}
\begin{quote}
The near-horizon geometry of parallel NS5-branes
is described by an exact superconformal 2d field theory based on
${\cal K}^{10}\equiv W_{k}^{4} \times{\cal M}^6$, with
${\cal M}^6$ a flat 5+1 space time and $W_{k}^{4} \equiv
U(1)\otimes SU(2)_{k}$, a four dimensional background with non-trivial
dilaton. The ten-dimensional ``BULK'' spectrum of excitations  of
${\cal K}^{10}$ can be derived  combining unitary representations
of the ${\cal N}=4$ superconformal theory of $ W_{k}^{4}$ and
${\cal M}^6$ in a modular-invariant  way. All Bulk states are massive
and belong to the long representations of the $N_6=2$ space-time
supersymmetry. The NS5-brane  states,   propagating  on  ${\cal M}^6$,
belong to the  short representations of  $N_6=2$. Both the bulk and
the brane spectrum is derived using the powerful worldsheet
technics of the  ${\cal N}=4$ superconformal theories. We claim that
both bulk and brane states are necessary to well define the theory.
The non abelian $U(n)$ or  $SO(2n)$ structure of the 5-brane
fields  follows from the  fusion coefficients appearing  in the
correlation functions involving   $SU(2)_k$ conformal fields.
\end{quote}
\vspace{.5cm}
\begin{flushleft}
\begin{center}
Talk given at the First workshop of the RTN network \\
{\it  ``The quantum structure of spacetime and the\\
 geometric nature of fundamental interactions''}\\ 
 and\\
 the 34th International Symposium Ahrenshoop on\\ 
{\it ``The Theory of Elementary Particles''}\\ 
 4-10 October 2000 
\vskip .3cm
\end{center}
\rule{8.1cm}{0.2mm}\\
$^{\star}$
{\small Research supported in part by
the EEC under  the contract
HPRN-CT-2000-00131,\\ ``the quantum structure of spacetime and
the  geometric nature of fundamental interactions''.} \\
$^{\dagger}$
{\small Unit\'e mixte du CNRS UMR 7644.} \\
{\small e-mail:  kounnas@physique.ens.fr}
\end{flushleft}
\end{titlepage}
\setcounter{footnote}{0}
\setcounter{page}{0}
\setlength{\baselineskip}{.7cm}
\newpage
%
% BODY
%
\section{Introduction}
The study of string propagation in non-trivial gravitational
backgrounds can provide a better understanding of quantum
gravitational phenomena. Non-trivial classical
string backgrounds can be obtained by two different methods.
The first makes use of a two-dimensional $\sigma$-model where
the space-time backgrounds correspond to field-dependent coupling
constants. The vanishing of the corresponding
$\beta$-functions is identified with the background field equations
of motion in the target space \cite{betafunct}. The second approach
consists of replacing the free space-time coordinates by non-trivial
(super) conformal systems, which, in the semiclassical limit, can be
interpreted as describing a string propagation in non-trivial
space-time. The two methods are useful and complementary.
The $\sigma$-model approach provides a clear geometric interpretation,
but it has the disadvantage of the $\alpha^{\prime}$-expansion, which
is valid only when all curvetures and derivatives on space-time fields
are small. The conformal field theory approach takes into account all
orders in $\alpha^{\prime}$ automatically and has the main advantage of
deriving exact string vacua. A  special class of exact solutions
are based on $N=4$ superconformal systems \cite{ademolo,kprn4,Kn=4,AFK}
where the  degrees of freedom of the ten supercoordinates form  three
superconformal systems
\be
\{ \hat c\} = 10 = \{ \hat c = 2 \} + \{ \hat c = 4 \}_1 + \{ \hat c
= 4\}_2~.
\ee

The $\hat c = 2$ two free superfields.  The time-like supercoordinate
and a  space-like supercoordinates.
The remaining eight supercoordinates appear
in  groups of four in $\{\hat c = 4\}_1$,  $\{\hat c = 4\}_2$~.
Both
$\{\hat c = 4\}_A$ systems exhibit $N = 4$ superconformal symmetry of
the Ademollo et al  \cite{ademolo,kprn4,Kn=4,AFK}. The non-triviality of
these solutions follows from the fact that there exist realizations of
the $\hat c = 4$, $N = 4$ superconformal systems that are based on
geometrical and topological non-trivial spaces \cite{kprn4,Kn=4,AFK},
other than the  $T^4/Z_2$ orbifold and the $K_3$ compact manifold.

 The first subclass is characterized by two
integer parameters $k_1$,
$k_2$, which are the levels of two $SU(2)$ group manifolds. For
weakly curved backgrounds (large $k_A$) these solutions can be interpreted
in terms of a ten-dimensional, but topologically non-trivial, target
space of the form  $R^4 \otimes S^3 \otimes S^3$.
In the special limit
$k_2\rightarrow \infty$ one obtains the semi-wormhole solution of
Callan, Harvey and Strominger \cite{wormclas}, based on a
six-dimensional flat background, combined with a four-dimensional space
$W_{k}^{(4)} \equiv U(1)\otimes SU(2)_{k}$.

 The underlying superconformal field theory
includes a supersymmetric $SU(2)_{k}$ WZW model
describing the three coordinates of $S^3$ as well as a non-compact
dimension with a background charge, describing the scale factor of
the sphere. $M^6\otimes W_{k}^{(4)}$ corresponds to the gravitational
{\it back reaction} of $n= k+2$  NS five-branes
\cite{wormclas, 5brane, fivebranes}.

A second subclass of solutions is based on
a different realization of
the $N=4$ superconformal system with $\hat c=4$,
it replaces the $W_k^{(4)}$ space by:
\be
\Delta_k^{(4)} \equiv \left\{\left[\frac{SU(2)}{U(1)}\right]_k
\otimes
 \left[ \frac{SL(2,R)}{U(1)}\right]_{k+4}\right\}_{\rm SUSY}~.
\ee
A  gauged supersymmetric WZW model with $\hat c[\Delta^{(4)}_k] =4$
for any  value of $k$ \cite{ Kn=4,AFK}.
The choice of the levels $k$ and $k+4$ is
necessary for the existence of an $N=4$ symmetry with $\hat c=4$.
Another subclass of solutions is obtained using the
{\it T- dual} \cite{RoVe,KLK} of $W_k^{(4)}$, $C^{(4)}_k$
\cite{ Kn=4,AFK}
\be
C^{(4)}_k \equiv \left(\frac{SU(2)}{U(1)}\right)_{k}\otimes
U(1)_{R}\otimes U(1)_{Q}
\ee
with a background charge $Q=\sqrt{2/(k+2)}$
in  one of the two coordinate currents ($U(1)_{Q}$).
The other free  coordinate ($U(1)_{R}$) is compactified
on a torus with radius
$R=\sqrt{2k}$.

\section{ 10-d Target Spaces,  via
$N=4$, ${\hat c}=4$ Conformal Block Construction \cite{ Kn=4,AFK}}

Having at our disposal non-trivial $N=4$, ${\hat c}=4$
superconformal systems, we can use them as building blocks to obtain
new classes of $exact$ and $stable$ string solutions around
non-trivial backgrounds in both type-II and heterotic superstrings.\\

$ {\cal M}^{2}\otimes W_{k_1} \otimes W_{k_2}~~~~~$
{\bf 9-branes or 2x5-branes}\\

$ {\cal M}^{6} \otimes W_{k}~~~~~$
{ \bf 5-branes,}\\

$ {\cal M}^{2}\otimes C_{k_1} \otimes C_{k_2}~~~~~$
{\bf T-duals of  9- or 2x5-branes}\\

$ {\cal M}^{6} \otimes C_{k}~~~~~$
{ \bf T-dual of 5-branes,}\\

$ {\cal M}^{2}\otimes C_{k_1} \otimes W_{k_2}~~~~~$
{\bf T-duals of 9- or  2x5-branes}\\

$ {\cal M}^{2}\otimes C_{k_1} \otimes \Delta_{k_2}~~~~~$
{\bf T-duals of   9- or 2x5-branes}\\

$ {\cal M}^{2}\otimes \Delta_{k_1} \otimes W_{k_2}~~~~~$
{\bf T-duals of   9- or 2x5-branes}\\

 $ {\cal M}^{2}\otimes \Delta_{k_1}
\otimes\Delta_{k_2}~~~~~$
{\bf T-duals of   9- or 2x5-branes}\\

${\cal M}^6 \otimes \Delta_k~~~~~$
{\bf T-dual of 5-branes,}\\

All the above superconformal systems correspond either  to
{\it  Non-critical Superstrings}\cite{kuseiberg,Kn=4,AFK}
and/or to a {\it Bulk 10d Superstring theory with branes}
\cite{wormclas, 5brane,Kn=4,AFK, fivebranes} or
to a {\it  background solution of the $SU(2)\times SU(2)$ Gauged
Supergravities}\cite{ABS,AFK}.

\section{The Spectrum of String Excitations in the Bulk}

The explicit realizations of the $N=4$ algebra in terms of
known conformal field theories,
allows us to compute the spectrum of string excitations in  any
of the background solutions given in the previous Section.

The $N=4$ superconformal symmetry implies
the existence of space-time supersymmetry \cite{banksdixon}in the
corresponding non-trivial target spaces,
and thus guarantees the stability of the  theory.

In all constructions, the total number of space-time supersymmetries
is reduced by a factor of 2 with respect to the flat (toroidal)
compactifications. In the context of the $\sigma$-model approach, the
$N=4$ spectral flows are related to the number of covariantly
constant spinors admitted by the corresponding non-trivial target
spaces \cite{wormclas}.

The reduction of space-time supersymmetries by a factor of 2
in $W^{4}$  non-trivial space is due to the
torsion and  background charge term appearing, in the
supercurrents \cite{AFK}. $Q^2={2}/{(k+2)}$

$$
T = -\frac{1}{2}\left[Q^2  J_i^2 + J_4^2
-\Psi_a\partial\Psi_a + Q\partial J_4\right]~,\nonumber\\
$$
$$
G_4 = Q\left( J_i\Psi_i+
\frac{1}{3}\epsilon_{ijl}\Psi_i\Psi_j\Psi_l+\partial\Psi_4 \right)
+ J_4\Psi_4 ~,\nonumber\\
$$
$$
G_i = Q\left(J_i\Psi_4-\epsilon_{ijl}J_j\Psi_l
+\epsilon_{ijl}\Psi_4\Psi_j\Psi_l-\partial\Psi_i \right)
-J_4\Psi_i ~,\nonumber\\
$$
\be
S_{i} = \frac{1}{2}\left(\Psi_{4}\Psi_{i}+
\frac{1}{2}\epsilon_{ijl}\Psi_{j}\Psi_{l}\right) ~.
\ee

The global existence of the (chiral) $N=4$ superconformal algebra
implies  a universal GSO projection that
generalizes the one of the the free-field realization
and it is responsible for the
existence of space-time supersymmetry.
The reduction of space-time supersymmetries by a factor of 2
implies an extra GSO projection involving the $SU(2)_k$
spin $j$ and the spins of the
 $SO(4)_{\Psi_I} \equiv SU(2)_+\times SU(2)_-$
constructed by the 2d fermions $\Psi_I, I=1,2,3,4$.

\subsection{The spectrum in a five-branes background}

The basic rules to construct the bulk  spectrum  are similar
to that of, the orbifold construction \cite{orbifold},
the free 2-d fermionic constructions \cite{abk4d},  and the
Gepner construction \cite{gepner4d}.
One combines in a modular invariant way the
world-sheet degrees of freedom consistently with unitarity and
spin-statistics of  the string spectrum.
The six non-compact coordinates,
together with the reparametrization ghosts ({\bf b},{\bf c}),
provide a contribution to the (type-II) partition function:

\be
Z_{B}[F^{(6)};({\bf b},{\bf c})]= \frac{{\rm Im}\tau^{-2}}{\eta ^{4}
(\tau)\bar{\eta}^{4}(\bar{\tau})}~.
\ee
 The contribution of the ${\cal M}^6$  world-sheet fermions together 
with the ${\bf \beta}$ and ${\bf \gamma}$ super-reparametrization 
ghosts is:
\be
Z_{F}[M^6;({\bf \beta},{\bf \gamma})]= (-)^{\alpha +\beta}~
\frac{\theta ^{2} (^{\alpha}_{\beta})}{ \eta ^{2}(\tau)}
~~ (-)^{\bar{\alpha} +\bar{\beta}}~
\frac{\bar{\theta}^{2}(^{\bar{\alpha}}_{\bar{\beta}})}
{\bar{\eta}^{2}(\bar{\tau})}~.
\ee

The Neveu-Schwarz ($NS$, $\overline{NS}$) sectors correspond to
$\alpha, \bar{\alpha}=0$ and the Ramond ($R,\bar{R}$) sectors
correspond to $\alpha, \bar{\alpha}=1$.

 Then, one must combine the above
$M^6$ characters with those of $W^{(4)}_k$ \cite{AFK}:\\

(i) the $SU(2)_k$ characters, ($\chi ^{L}_k~,~L=1,2,\cdots,k$),\\

(ii) the $U(1)_Q$ Liouville characters,\\

(iii) the $SU(2)_+$,  ($\chi ^{l}_+~,~ l=0,1 $),\\

(iv) the $SU(2)_-$,  ($\chi ^{l}_-~,~ l=0,1 $).\\
$~$\\

$\bullet$ {The $U(1)_Q$ Liouville characters:}\\

i){\it Continuous Representations}
generated by the lowest-weight operators:
\be
e^{\beta X_L}\ ;\quad \beta=-\frac{1}{2}Q +ip\ ,
\ee
with {\it positive conformal weights}\\

$$
h_p=\frac{Q^2}{8}+\frac{p^2}{2}.
$$

The fixed imaginary part in the momentum $iQ/2$ of the plane waves,
is due to the non-trivial dilaton motion.\\

ii){\it Discrete Representations}.
They  correspond to lowest-weight operators
 with $\beta=Q\tilde{\beta}$ real, leading to
{\it negative conformal weights}
\be
h=-\frac{1}{2}\tilde{\beta}(\tilde{\beta}+1)Q^2=
-\frac{\tilde{\beta}(\tilde{\beta}+1)}{k+2}
\ee

Both categories of Liouville representations give rise to
{\it unitary representations} of
the $N=4$, ${\hat c}=4$ system $W^{(4)}_k$, once they are combined
with the remaining degrees of freedom.

{\it The  continuous representations}
form {\it long} (massive) representations
of $N=4$ with conformal weights larger than the $SU(2)$ spin, { $h>S$}
\cite{ademolo,AFK}.
All bulk states belong to long representations of the $N=4$.

One the other hand the discrete representations form
{\it short} representations of
$N=4$ with {$h=S$}, \cite{ademolo,AFK} and  thus { $\beta$}
take only a {\it finite number of values} \cite{AFK}\\
\be
-(k+2)/2~\le~\tilde{\beta}~\le~ k/2 ~.
\ee

In fact, their locality with
respect to the
$N=4$  operators implies:
$$
S = \frac{1}{2},\quad \tilde{S}=\frac{1}{2}:\quad \tilde{\beta} =
-(j+1) ~,\nonumber \\
$$
\be
S = 0,\quad \tilde{S}=0:\quad \tilde{\beta} = j ~,
\ee
The vertex operators corresponding to the discrete representations
take the form
\be
R[~S,j,{\tilde \beta}~]= V_{\Psi}(S)~{\bar V}_{\bar{\psi}}({\bar S})
~\chi_k(j)~ {\bar\chi}_k(j)~ e^{\beta X_L} ~,
\ee
have fixed  conformal weights in the -1 picture,
$(h_L, h_R)=(\frac{1}{2},\frac{1}{2})$, giving rise to massless states
belonging to the short representations of the $N=4$ \cite{AFK}.
They correspond to the  5-Brane States propagating on ${\cal M}^6$
\cite{fivebranes}. In the  massless
spectrum there are either  $U(k+2)$  or $SO(k+4)$ \cite{fivebranes}.
In the first case 2j integer ($SU(2)_k$ 2d WZW-model) while in the
second case,  j takes integer values  ($SO(3)_{k/2}$ 2d WZW-model).

 All Bulk States are massive with minimal mass
$m^2_{\rm min}= Q^2/4$ due to the linear dilaton \cite{ABEN}
and torsion \cite{AFK, infrared}.
A modular invariant partition function for the BULK states
($k$ even)  is \cite{AFK}:\\

$$
Z_W = {{\rm Im}~\tau^{-5/2}\over \eta^5{\bar\eta}^5}~~
{1\over 8}
\sum_{\alpha,\beta,{\bar\alpha},{\bar\beta},\gamma,\delta}~~
(-)^{\alpha+\beta}~
\frac{\theta^{2}(^{\alpha}_{\beta})}{\eta^2}
\frac{\theta^{2}(^{\alpha+\gamma}_{\beta+\delta})}{\eta^2}~\times
$$

\be
~~~~~(-)^{{\bar\alpha}+{\bar\beta}}~
\frac{{\bar\theta}^{2}(^{\bar\alpha}_{\bar\beta})}{{\bar\eta}^2}
\frac{{\bar\theta}^{2}
(^{{\bar\alpha}+\gamma}_{{\bar\beta}+\delta})}{{\bar\eta}^2}
~~~(-)^{\delta(\alpha+{\bar\alpha}+{k\over 2}\gamma)}~
Z_k[^{\gamma}_{\delta}]
\ee

The first factor is  the contribution of
the non-compact coordinates and that
of the Liouville mode.

$Z_k[^{\alpha}_{\beta}]$
defines appropriate character combinations of
$SU(2)_k$. Under modular transformations \cite{AFK}:

$$
Z_k[^{\alpha}_{\beta}] = \sum_{L=0}^k e^{i\pi\beta L}~ \chi_k^L
~\bar{\chi}_k^{L+\alpha (k-2L)} \ ,
$$
$\alpha$, $\beta$ can be either 0 or 1.
$$
\tau\rightarrow\tau +1~~~:~~~
Z_k[^{\alpha}_{\beta}] \longrightarrow
e^{-i\pi\frac{k}{2}\alpha^2}~ Z_k[^{~\alpha}_{\beta+\alpha}]
$$
$$
\tau\rightarrow{-1/\tau}~~~:~~~
Z_k[^{\alpha}_{\beta}]\longrightarrow
e^{i\pi k\alpha\beta}~ Z_k[^{\beta}_{\alpha}]\ .
$$
the $\beta$ and
${\bar\beta}$ summations give rise to  universal (left- and
right-moving) GSO projections, which imply the existence of
space-time supersymmetry.
The summation over $\delta$ gives rise to an
additional projection, which correlates the $SU(2)_-$ (left and
right) spin together with the spin of $SU(2)_k$ . It  reduces
the number of  supersymmetries by a factor of 2.
\be
2{\tilde S}_2 + 2{\bar{\tilde S}}_2 + L + {k\over 2}\gamma = {\rm
even}.
\ee
In the $\gamma=0$ sector, the lower-lying states have (left and
right)
mass-squared $Q^2/8$ and $L=0$. It is
convenient to classify the states in the context of a
six-dimensional theory.
The lower-lying states come from the
gravitational supermultiplet of the six-dimensional $N=2$
supergravity:
$$
(|\Psi^{\mu}> + |({\rm spin}\Psi^{\mu})_- ({\rm spin}\Psi^{I})_+>)
\otimes
(|{\bar\Psi}^{\mu}> + |({\rm spin} {\bar\Psi}^{\mu})_-
({\rm spin}{\bar\Psi}^{I})_+>)
~e^{ip_{\mu}X^{\mu}+(ip -{Q\over 2})X_L}
$$
together with four vector multiplets:
$$
(|\Psi^{I}> + |({\rm spin}\Psi^{\mu})_+ ({\rm spin}\Psi^{I})_->)
\otimes
 (|{\bar\Psi}^{I}> + |({\rm spin}{\bar\Psi}^{I})_+
({\rm spin}{\bar\Psi}^{I})_->)
~e^{ip_{\mu}X^{\mu}+(ip -{Q\over 2})X_L}~.
$$
As expected from the effective field theory point of view,
their mass-squared $Q^2/8$ is due to the dilaton motion for bosons,
and to the non-trivial torsion for fermions.\\

 The $(\gamma=1)$ (Twisted) sector contains states
with (left and right) mass-squared always larger than $(k-2)/16$
For any $k>2$, the twisted states have masses larger than $Q^2/8$
and the lower mass spectrum comes always from the $L=0$ states
contained in the untwisted sector.
In that sense $k=2$ is an
exceptional case \cite{AFK}, since the lower-lying twisted states
are massless with $L=\bar{L}=k/2=1$. These states form massless unitary
representations (short) of the $N=4$ $\hat{c}=8$.

All other {\it short} multiplets are coming from the n-NS5-Branes
propagating in the six dimensional target space.

\be
R[~S,j,{\tilde \beta}~]~e^{ik_{\mu}X^{\mu}}, ~~~~~{\rm with}~~~~
(h_L,h_R)=(\frac{1}{2},\frac{1}{2})~.
\ee

\section{ Conclusions}

Superstring  solutions in the semiclassical limit define
background solutions of the  extended supergravities. This limit turns
out to be very useful regarding the study of the  string-induced
low-energy theories, as well as the study of  physics in weakly
curved domains of space time.
The  supergravity field theory picture  fails when
the involved curvetures are strong. It is then
necessary  to go beyond the semiclassical limit
and work directly on the string level, using the powerful techniques
of the underlying two-dimensional (super) conformal field theory.

For a generic string background the stringy approach is at present
non-accessible.  It is possible to go further in the stringy direction
for some special backgrounds based on the $N=4$  superconformal symmetry.
Some of the constructions are connected to the
non-critical strings and to some stable solutions of the  gauged
supergravities.

 The ten-dimensional ``bulk'' spectrum of excitations
can be derived  combining unitary representations of the $N=4$
superconformal theory in a modular-invariant way.
In the case of $W_{k}^{4}$ constructions, these representations
are expressed in terms of the well-known $SU(2)$ characters,
while in all other constructions one uses also  the characters  of
some compact $SU(2)/U(1)$)  and/or non-compact $SL(2,R)/U(1)$
parafermions \cite{gepner4d, DLPnoncompact}.
In the five-brane construction all bulk  states are massive
as soon as  $k$ is large (Massive Representation of N=4).
When $k=2$, there extra massless ``twisted'' states other
than the Five-brane States  ($U(n)$ or $SO(2n)$ vector multiplet states).

The  Non-propagating brane states correspond to Liouville
Discrete representations  with negative conformal weight.
The non abelian $U(n)$ and $SO(2n)$ structure of the 5-brane
fields,  $<R(j_1) R(j_2) R(j_3)>\sim F_{j_1,j_2,j_3}$, follows from the
correlation of the  $SU(2)_k$ vertex operators \cite{fivebranes}.

\vskip0.5cm
\noindent
{\large \bf Acknowledgements}
This work is partially supported by the EEC under the contract
HPRN-CT-2000-00131, ``the quantum structure of spacetime and
the  geometric nature of fundamental interactions''.
\smallskip
\noindent

%%%%%%%%%%%%%%%%%%%%%%
%%%%%%%%%%%%%%%%%%%%%%

\end{document}